\documentclass[prd,twocolumn,10pt]{revtex4-1}
\usepackage{graphicx,color}
\usepackage{amsmath,amsfonts,amssymb}

\begin{document}

\title{Spectral dimension with deformed spacetime signature}

\author{Jakub Mielczarek$^{*}$ and Tomasz Trze\'{s}niewski$^{*,\dagger}$ \\
${}^*$Institute of Physics, Jagiellonian University, ul.\ {\L}ojasiewicza 11, 
30-348 Krak\'{o}w, Poland \\
${}^{\dagger}$Institute for Theoretical Physics, University of Wroc\l{}aw, pl.\ 
Borna 9, 50-204 Wroc\l{}aw, Poland}

\date{\today}

\begin{abstract}

Studies of the effective regime of loop quantum gravity (LQG) revealed that, in the limit of Planckian 
curvature scales, spacetime may undergo a transition from the Lorentzian to Euclidean signature. 
This effect is a consequence of quantum modifications of the hypersurface deformation algebra, which 
in the linearized case is equivalent to a deformed version of the Poincar\'{e} algebra. In this paper the 
latter relation is explored for the LQG-inspired hypersurface deformation algebra that is characterized 
by the above mentioned signature change. 

While the exact form of the deformed Poincar\'{e} algebra is not uniquely determined, the algebra 
under consideration is representative enough to capture a number of qualitative features. In particular, 
the analysis reveals that the signature change can be associated with two symmetric invariant energy 
scales, which separate three physically disconnected momentum subspaces. 

Furthermore, the invariant measure on momentum space is derived, which allows to properly 
define the average return probability, characterizing a fictitious diffusion process on spacetime. 
The diffusion is subsequently studied in the momentum representation for all possible variants 
of the model. Finally, the spectral dimension of spacetime is calculated in each case as a function 
of the scale parameter. In the most interesting situation the deformation is of the asymptotically 
ultralocal type and the spectral dimension reduces to $d_S = 1$ in the UV limit. 
\end{abstract}

\maketitle

\section{Introduction} 

Analysis of a (fictitious) diffusion process on space or spacetime has become a versatile tool for 
characterizing classical and quantum models. The usefulness of this method stems from the spectral 
properties of Laplace operators, which allow us to determine such quantities as the return probability 
or spectral dimension. The former can be applied to calculations of the vacuum energy density and 
entanglement entropy \cite{Nesterov:2011gy}. Studies of the spectral dimension, which is one of the 
possible tools allowing to characterize the spacetime dimensionality at different scales, has become 
especially popular over the last ten years in the context of quantum gravity. In particular, this is so 
because the spectral dimension can tell us what is effective spacetime dimension perceived by a field 
defined of a given spacetime. The results may also be used for a comparison of different models of 
the Planck scale physics. However, the potential convergence of their predictions has to be treated 
as a hint rather than indication of the more fundamental relation. The reason is that different structures 
of (quantum) geometry may lead to similar running of the spectral dimension with scale. 

The above mentioned avalanche of papers concerning the spectral dimension has been triggered 
by the seminal analysis performed in the framework of Causal Dynamical Triangulations (CDT) 
\cite{Ambjorn:2005db}. In the space of parameters of this model one can distinguish several 
phases, characterized by different geometrical properties. Deep within phase C, in which one obtains 
an extended physical universe, it has been shown that spacetime undergoes a dimensional reduction 
from the classical value $d_S \approx 4$ at large diffusion times (IR limit) to $d_S \approx 2$ at small 
times (UV limit). The result has been subsequently generalized, showing that the UV value of the spectral 
dimension varies depending on a position on the CDT phase  diagram. In particular, in the region of 
the phase C close to the phase A (where spacetime becomes a sequence of short-living small 
universes) the value $d_S \approx 3/2$ has been measured, which, interestingly, can serve as a 
resolution to the entropic argument against the asymptotic safety scenario \cite{Coumbe:2014noa}. 

The dimensional reduction to $d_{UV} = 2$ in the ultraviolet limit seems to be a common feature 
of a variety of approaches to quantum gravity \cite{Carlip:2009kf}. It has been observed in (besides 
CDT): Ho\v{r}ava-Lifshitz gravity (for the characteristic exponent $z = 3$) \cite{Horava:2009st}, 
asymptotic safety scenario \cite{Lauscher:2005fy}, multi-fractal spacetimes \cite{Calcagni:2012qn}, 
causal sets (probed by a scalar field) \cite{Carlip:2016dy,Belenchia:2016ss} and spin-foam models 
\cite{Modesto:2008jz,Calcagni:2014cza}. On the other hand, the value $d_{UV} = 2$ is not completely 
universal but typical for a certain class of models, describing a specific type of quantum spacetime 
configurations, perhaps a specific phase of the gravitational field. In particular, as we already 
mentioned, the values different from $2$ can be found on the CDT phase diagram. While 
in the physical phase C the dimensional reduction from $d_{IR} = 4$ to $d_{UV} \in [3/2, 2]$ has 
been observed, in two other phases values of the spectral dimension are significantly different, e.g.\! 
in the non-geometric phase B (where universe is a highly-connected single time slice) the spectral dimension 
diverges at small scales. Another example of $d_{UV} \neq 2$ is provided by the $\kappa$-Minkowski 
noncommutative spacetime, which is often employed in models of doubly special relativity or relative 
locality, characterized by deformed relativistic symmetries. It has been shown to exhibit the dimensional 
reduction from $d_{IR} = 4$ to $d_{UV} = 3$ \cite{Benedetti:2008gu} but this value is obtained for a 
particular choice of the Laplace operator on momentum space. Meanwhile, in the case of the relative 
locality-inspired Laplacian the spectral dimension diverges at small scales \cite{Arzano:2014jfa}. One 
of the issues in the above context is also the interplay between the running dimension and either 
breaking or deforming of the relativistic symmetries of spacetime \cite{Amelino:2014pe}. 

In this paper we extend previous results by investigating the diffusion on spacetime whose 
Poincar\'{e} symmetries are deformed by effects predicted in the effective regime 
of loop quantum gravity (LQG). Namely, it has been shown that as a consequence of 
requiring the anomaly freedom the classical algebra of gravitational constraints (the 
hypersurface deformation algebra) is subject to a quantum deformation \cite{Bojowald:2011aa,Cailleteau:2011kr}. 
A pronounced manifestation of the deformation is the phenomenon of a``dynamical'' signature change 
\cite{Cailleteau:2011kr,Mielczarek:2012pf,Bojowald:2015gra}. Here, in particular, we study the role of such a signature change in 
the description of a diffusion process. Our calculations are performed in the symmetry-reduced setup, where the 
isotropy and homogeneity of spacetime are assumed. Then the hypersurface deformation algebra 
is equivalent to a deformed version of the Poincar\'{e} algebra. The explicit form of the latter deformation has so far 
been obtained only in the spherically symmetric case \cite{Bojowald:2012ux}, where the 
algebra is effectively two-dimensional. The corresponding deformed 
Poincar\'{e} algebra (and, tentatively, the coalgebra) was also studied in \cite{Amelino-Camelia:2016gfx,
Brahma:2016tsq,Ronco:2016rtp}. An additional effect of such investigations is that they provide a desired 
prediction of the deformed relativistic symmetries from the hypothetical full theory of quantum gravity, while 
usually they have only been introduced a priori, in the phenomenological approach \cite{Amelino:1999ds}. 
Finally, let us mention that in \cite{Ronco:2016rtp} it has been suggested that the loop deformation 
of the hypersurface deformation algebra may lead to the dimensional reduction to $d_{UV} = 2.5$ 
near the Planck scale. However, this value can change when higher order contributions to the dispersion 
relation are taken into account. 

In the next sections, inspired by the analysis performed in \cite{Mielczarek:2013rva}, we recover the form of 
a particular deformed four-dimensional Poincar\'{e} algebra by imposing certain reasonable conditions. 
While such a method does not lead to the unique form of the deformation, the obtained results allow to 
deduce a number of qualitative conclusions. In particular, as we show, the deformation may lead to the 
appearance of a new invariant energy scale. Furthermore, we explicitly derive the invariant measure on 
momentum space, which turns out to be different from the one that could be naively expected from the 
form of the mass Casimir. The symmetry algebra turns out to be compatible with the standard Heisenberg 
algebra of phase space variables and therefore we assume that the latter algebra is undeformed. Finally, 
(in the momentum representation) we calculate the average return probability and the resulting spectral 
dimension of spacetime, which is a function of the scale parameter. We analyze four distinct cases, 
depending on the deformation parameter. The physical meaning and relations of the results with other 
approaches are discussed.

Throughout this paper we use the Planck units, where $\hbar = c = 1$, $G = 1/m^2_{\text{Pl}}$ 
and $m_{\text{Pl}}$ denotes the Planck mass. 

\section{Deformed Poincar\'{e} algebra} \label{DefPoinc}

The perturbative analysis of the effective regime of loop quantum gravity has revealed 
(see e.g. \cite{Bojowald:2011aa,Cailleteau:2011kr,Bojowald:2015gra}) the 
following form of the quantum-deformed hypersurface deformation algebra:
\begin{align}
\left\{ D[N^a_1],D[N^a_2] \right\} &= D \left[ N_1^b \partial_b N^a_2 - N_2^b \partial_b N^a_1 \right], \label{DHDA1}  \\
\left\{ S[N],D[N^a] \right\} &= -S \left[ N^b \partial_b N \right], \label{DHDA2}  \\
\left\{ S[N_1],S[N_2] \right\} &= D \left[ s \Omega q^{ab} (N_1 \partial_b N_2 - N_2 \partial_b N_1) \right], \label{DHDA3}
\end{align}
where $q^{ab}$ denotes the spatial metric, with spatial indices $a,b = 1,2,3$, and $s$ is the 
spacetime metric signature. 
$D[N^a]$ is the constraint generating spatial diffeomorphisms, parametrized by a shift vector field $N^a$, 
while $S[N]$ is the scalar constraint and generator of deformations in the direction normal to spatial hypersurfaces, 
parametrized by a lapse function $N$. Due to the effects of LQG the scalar constraint is subject to a quantum 
deformation. Moreover, the hypersurface deformation algebra itself is deformed through the presence of the 
factor $\Omega$, which is some function of gravitational field variables. The general form of this function in LQG 
is not yet known but it has been explicitly derived for specific symmetry reduced configurations. In the case 
when the sign of $\Omega$ is constant the deformation can actually be absorbed by an appropriate transformation of variables, 
which cannot be achieved if the sign is changing \cite{Bojowald:2016hgh}. 

In particular, in the case of a homogeneous and isotropic spacetime configuration on which there are 
introduced perturbative inhomogeneities with holonomy corrections the deformation factor is given by \cite{Cailleteau:2011kr}
\begin{equation}
\Omega = \cos(2\gamma \bar{\mu} \bar{k}) \cong 1 - 2 \frac{\rho}{\rho_c} \in [-1,1]\,. \label{eq:II.01}
\end{equation}
Here $\cong$ denotes imposition 
of the constraint $S[N]$, $\bar{p}$ and $\bar{k}$ are the homogeneous Ashtekar variables, $\gamma \sim 1$
is the Barbero-Immirzi parameter and $\bar{\mu} = \sqrt{\Delta/\bar{p}}$ the lattice refinement, with $\Delta$ being 
the minimal area, expected to be of the order of the Planck area $\Delta \sim 1/m_{\text{Pl}}^2$. $\rho$ is energy density of the matter content of universe and $\rho_c = 3m_{\text{Pl}}^2/(8\pi \gamma \Delta) \sim 
m_{\text{Pl}}^4 $ the maximal allowed value of energy density, expected to be of the order of the Planck density. 
We note that the classical value $\Omega = 1$ is correctly recovered in the limit of low energy densities $\rho \rightarrow 0$. 
On the other hand, when the density reaches the maximum $\rho \rightarrow \rho_c$, 
the opposite value $\Omega = -1$ is achieved. 

Since the deformation factor $\Omega$ in (\ref{DHDA3}) is accompanied by the metric signature $s$, 
the change of the sign of $\Omega$ can be interpreted as the signature change \cite{Mielczarek:2012pf}. Namely, at low energy densities we have $s \Omega \rightarrow s$, which corresponds to the classical Lorentzian (for $s = 1$) or Euclidean (for $s = -1$) space, while at the maximal density $\rho = \rho_c$, in the deep quantum regime, the effective signature becomes $s \Omega \rightarrow -s$. Moreover, at $\rho = \rho_c/2$ the sign of $s \Omega$ turns out to be indefinite 
(i.e.\! $s \Omega = 0$), which can be associated with the state of so-called asymptotic silence or ultralocality 
\cite{Mielczarek:2012tn}. 

In this paper we are going to consider the algebra (\ref{DHDA1}-\ref{DHDA3}) restricted to linear 
hypersurface deformations, characterizing homogeneous and isotropic spacetime configurations. 
In such a case the above deformed algebra of constraints (for $s = 1$) is equivalent to a certain deformation of the 
Poincar\'{e} algebra, by which we mean the universal enveloping algebra of a Lie algebra $\mathfrak{iso}(3,1)$. 

As we discuss it in Appendix, the limit of linear deformations can be imposed on (\ref{DHDA1}-\ref{DHDA3}) by choosing the spatial metric to be given by the Kronecker delta, $q_{ab} = \delta_{ab}$ and restricting the 
expressions for a lapse function and shift vector to the linear form:
\begin{equation}
N(x) = \Delta t + v_a x^a\,, \quad N^a(x) = \Delta x^a + {R^a}_b x^b\,, \label{NNlin}
\end{equation}
where $\Delta t$, 
$\Delta x^a$, $v_a$ and ${R^a}_b$ are the infinitesimal parameters of, respectively, a time translation, spatial translations, boosts and rotations. The rotation matrix can be expressed in terms of infinitesimal angles $\varphi^a$ 
as $R^{ab} = \epsilon^{bac} \varphi_c$. Under such assumptions the scalar and diffeomorphism constraints can 
be expanded into the following combinations of the Poincar\'{e} algebra generators:
\begin{align}
S[N] &= S[\Delta t + v_a x^a] = -\Delta t P_0 - v^a K_a\,, \label{Slin}\\
D[N^a] &= D[\Delta x^a + {R^a}_b x^b] = -\Delta x^b P_b - \varphi^b J_b\,. \label{Dlin}
\end{align}
where $P_0$, $P_a$, $J_a$ and $K_a$ denote the generators of time translations, spatial translations, boosts and rotations, respectively.

In Appendix we show how the (standard) Poincar\'{e} algebra can be recovered by imposing the conditions $q_{ab} = \delta_{ab}$ and (\ref{Slin}-\ref{Dlin}) 
on the classical hypersurface algebra. The loop deformed algebra 
(\ref{DHDA1}-\ref{DHDA3}), although very similar to the classical one, does not allow to straightforwardly apply the same methodology. 
While for the undeformed brackets (\ref{DHDA1}-\ref{DHDA2}) the derivation of the 
corresponding sector of the Poincar\'{e} algebra is the same as in Appendix, the 
bracket (\ref{DHDA3}) contains the additional function $\Omega$ inside the 
diffeomorphism constraint. The most convenient solution would be to find a way to extract $\Omega$ in front of the 
constraint. However, apart from the perturbative approach, which has been used e.g.\! in deriving 
(\ref{eq:II.01}), it is not possible to do so directly. 

The strategy that we are going to apply here is based on the observation that a diffeomorphism 
constraint with some additional function $f$ of field variables inside can always be rewritten as the unmodified diffeomorphism constraint multiplied by a certain functional of $f$. Namely,
\begin{equation}
D[f N^a] = F[f] D[N^a]\,,
\end{equation}
where the functional $F[f]$ is simply given by
\begin{equation}
F[f] = \frac{D[f N^a]}{D[N^a]} =: \langle f \rangle_D\,,
\end{equation}
which can be interpreted as the diffeomorphism average of the function $f$. Then 
the problematic right hand side of (\ref{DHDA3}) can be expressed as
\begin{align}
& D\left[ s \Omega q^{ab} (N_1 \partial_b N_2 - N_2 \partial_b N_1) \right]  \nonumber\\
&= s \langle \Omega \rangle_D D\left[q^{ab} (N_1 \partial_b N_2 - N_2 \partial_b N_1) \right] \nonumber\\
&= -s_{\text{eff}}(v^a P_a + \varphi^a J_a)\,, \label{Dexpseff}
\end{align}
where we introduce the effective metric signature 
\begin{align}
s_{\text{eff}} = s \tilde{\Omega} &:= s \langle \Omega \rangle_D \nonumber\\
&= s \frac{D\left[ \Omega q^{ab} (N_1 \partial_b N_2 - N_2 \partial_b N_1) \right]}{D\left[ q^{ab} 
(N_1 \partial_b N_2 - N_2 \partial_b N_1) \right]}\,. \label{eq:II.02}
\end{align}
The quantity $s_{\text{eff}}$ has been called ``effective signature'' since it reduces to 
the metric signature $s$ in the classical limit, while otherwise it is instead a certain function of the 
Poincar\'{e} algebra generators. Because $s_{\text{eff}}$ multiplies the $P_a$ and $J_a$ 
generators in (\ref{Dexpseff}) in the same way as signature $s$ in the classical case, 
it is possible to expand the bracket (\ref{DHDA3}) analogously to its classical 
counterpart (\ref{HDA3}). Consequently, using the procedure discussed in Appendix 
we obtain the deformed Poincar\'{e} algebra that is determined by the following brackets: 
\begin{align}
\left\{J_a,J_b \right\} &= \epsilon_{abc} J^c\,, \label{DP1}\\
\left\{J_a,K_b \right\} &= \epsilon_{abc} K^c\,, \\
\left\{K_a,K_b \right\} &= -s_{\text{eff}} \epsilon_{abc} J^c\,, \label{DP3}\\
\left\{J_a,P_b \right\} &= \epsilon_{abc} P^c\,, \\
\left\{J_a,P_0 \right\} &= 0\,, \\
\left\{K_a,P_b \right\} &= \delta_{ab} P_0\,, \\
\left\{K_a,P_0 \right\} &= s_{\text{eff}} P_a\,, \label{DP7}\\
\left\{P_a,P_b \right\} &= 0\,, \\
\left\{P_a,P_0 \right\} &= 0\,, \label{DP9}
\end{align}
where the effective signature $s_{\text{eff}} = s \tilde{\Omega}$ appears only where $s$ is located in the standard 
Poincar\'{e} algebra. As one can see, $\tilde{\Omega}$ is some unknown function of the symmetry generators, 
related via (\ref{eq:II.02}) to the deformation factor $\Omega$, which is present at the level of the hypersurface 
deformation algebra. If the form of $\Omega$ was known as a function of field variables, then one could also try 
to derive the form of $\tilde{\Omega}$, using e.g.\! the Brown-York generators \cite{Brown:1992br}. 

Precisely speaking, our deformed Poincar\'{e} algebra is 
a certain quotient of the tensor algebra defined by the brackets (\ref{DP1}-\ref{DP9}) and the latter is actually 
a semi-classical deformation of a Poisson algebra. In other words, $P_0$, $P_a$, $J_a$ and $K_a$ can be 
viewed as functions on classical phase space (spacetime and the corresponding momentum space) but satisfying 
the deformed brackets. Passing to the quantum theory would require replacing $\left\{\cdot,\cdot\right\} 
\longrightarrow \frac{1}{i} [\cdot,\cdot]$ and promoting $P_0$, $P_a$, $J_a$ and $K_a$ to quantum operators. 
However, for the purposes of this paper it is not necessary to discuss the full quantum picture 
(in particular, in the framework of quantum groups) and therefore we restrict the considerations to the 
semi-classical regime. 

As mentioned above, a priori we do not know the functional form of the deformation factor $\tilde{\Omega}$. However, 
we are able to constrain it using 
mathematical consistency and experience gained from the symmetry reduced models, as well as calculate it explicitly in some particular cases. Let us first note that since $\tilde{\Omega}$ is generally a non-trivial function of the algebra generators, the brackets 
(\ref{DP1}-\ref{DP9}) do not define a Lie algebra. Nevertheless, it is reasonable to require the Jacobi identities, which are a feature of the 
Poincar\'{e} (Lie) algebra, to be satisfied even for a deformed version of this algebra. 
The main motivation to do so is that in such a case the (standard) Leibnitz rule for the action of an algebra on itself is preserved. One can also naturally expect that $\tilde{\Omega}$ is rotationally invariant. Moreover, the ordinary 
Poincar\'{e} algebra should be recovered in the limit of vanishing deformation (which will correspond to low energies). Our final 
assumption, which we choose here as a specific simple example, is that $\tilde{\Omega}$ depends only on the translation generators and is a separable function of the form 
$\tilde{\Omega} = \tilde{\Omega}(P_0,{\bf P}^2) = A(P_0) B({\bf P}^2)$, where $A$ and $B$ can be determined from the 
other conditions. If we allowed $\tilde{\Omega}$ to depend on rotations $J_a$ and boosts $K_a$, the analysis and 
interpretation of results would be much more difficult. 

In the case of Lorentzian signature $s = 1$ the Jacobi identities lead to a differential equation 
on $\tilde{\Omega} = \tilde{\Omega}(P_0,{\bf P}^2)$, which has the solution 
\begin{equation}
\tilde{\Omega}(P_0,{\bf P}^2) = \frac{P^2_0/c_1 - 1}{{\bf P}^2/c_2 - 1}\,, \label{OmegaSolution}
\end{equation}
with two independent real constants $c_1$, $c_2$. As we will now show, the classical limit imposes a 
relation between $c_1$ and $c_2$. 

By construction, (\ref{OmegaSolution}) commutes with all generators of the algebra (\ref{DP1}-\ref{DP9}) 
and hence it is a Casimir element of the latter. Then the mass Casimir element $\mathcal{C}_1$ of the algebra 
(determining the mass square of a particle whose symmetries are described by (\ref{DP1}-\ref{DP9})), 
with the standard classical limit, can be constructed as the appropriate combination of $\tilde{\Omega}$ and the 
unit element, namely
\begin{equation}
\mathcal{C}_1 = a_1 \tilde{\Omega} + a_2 \mathbb{I} = \frac{-\frac{a_1}{c_1} P_0^2 
- \frac{a_2}{c_2} {\bf P}^2 + (a_1 + a_2)}{1 - c_2^{-1} {\bf P}^2}\,.
\end{equation}
The conditions that have to be satisfied in order to recover the proper classical limit are 
$a_2 = -a_1$, $a_1 = c_1$ and $c_1 = c_2 \equiv \alpha$, where $\alpha$ is a (positive 
or negative) real constant. Consequently, the mass Casimir can be written 
as \cite{Mielczarek:2013rva}
\begin{equation}
\mathcal{C}_1 = \frac{-P_0^2 + {\bf P}^2}{1 - \alpha^{-1} {\bf P}^2} \label{CasimirLorentz}
\end{equation}
and the classical expression is recovered for $|\alpha| \rightarrow \infty$.

Furthermore, taking into account the fact that $c_1 = c_2 \equiv \alpha$, the deformation factor (\ref{OmegaSolution}) 
simplifies to 
\begin{equation}
\tilde{\Omega}(P_0,{\bf P}^2) = \frac{P^2_0 - \alpha}{{\bf P}^2 - \alpha}\,. \label{OmegaLorentz}
\end{equation}
Let us note that, in contrast to (\ref{eq:II.01}), $\tilde{\Omega}$ is not a bounded quantity. In the 
calculations we assumed that in the limit $|\alpha| \rightarrow \infty$ the expression (\ref{OmegaLorentz}) 
tends to $1$, restoring the undeformed Poincar\'{e} algebra. Below we will extend the symmetry algebra 
by the Heisenberg algebra and identify the translation generators with components of a momentum. 
Then (\ref{OmegaLorentz}) can also be seen as tending to $1$ in the low energy limit $P_0 \rightarrow 0$, ${\bf P} \rightarrow 0$. 

Let us now consider an extension of the symmetry algebra (\ref{DP1}-\ref{DP9}) by the (undeformed) Heisenberg algebra of spacetime positions $x_\mu$ and four-momenta $p_\mu$, defined by the brackets
\begin{equation}
\left\{x_\mu,x_\nu\right\}  = 0\,, \quad \left\{x_\mu,p_\nu\right\}  =  \eta_{\mu\nu}\,, \quad \left\{p_\mu,p_\nu\right\} = 0\,, \label{UH}
\end{equation}
with spacetime indices $\mu,\nu = 0,1,2,3$ and where  $\eta = {\rm diag}(-1,1,1,1)$ is the Minkowski metric. Applying the approach of \cite{Kovacevic:2012an} we find the following realization of (\ref{DP1}-\ref{DP9}) in terms of the $x_\mu$ and $p_\mu$ generators:
\begin{align}
\epsilon_{abc} J^c &= x_a p_b - x_b p_a\,, \\
K_a &= x_a p_0 - x_0 p_a \tilde{\Omega}\,, \\
P_a &= p_a\,, \qquad P_0 = p_0\,. \label{eq:II.03}
\end{align}
The above expressions do not depend on the explicit form of $\tilde{\Omega} = \tilde{\Omega}(P_0,P_a)$. Then for $\tilde{\Omega}$ given by (\ref{OmegaLorentz}) we find that the brackets of $x_\mu$ with the boost generators have the non-trivial form
\begin{align}
\left\{K_a,x_0\right\} &= x_a - 2x_0 \frac{p_0 p_a}{p_0^2 - \alpha} \tilde{\Omega}\,, \\
\left\{K_a,x_b\right\} &= x_0 \tilde{\Omega} \left(\delta_{ab} - 2\frac{p_a p_b}{{\bf p}^2 - \alpha}\right)
\end{align}
and satisfy all necessary Jacobi identities. In this sense our deformed Poincar\'{e} 
algebra is compatible with the standard Heisenberg algebra (\ref{UH}). The merger of these two algebras describes usual commutative spacetime and corresponding momentum space but endowed with 
deformed relativistic symmetries. However, such an extension of the symmetry algebra is ambiguous, since the latter alone is actually insufficient to determine the form of the Heisenberg algebra. To this end we 
need also the coproduct and antipode on the algebra (\ref{DP1}-\ref{DP9}), which would turn the latter 
into a Hopf algebra. Then the appropriate Heisenberg algebra could generally be constructed using the 
so-called smash product construction, see e.g.\! \cite{Borowiec:2010ks}. Therefore, in principle, 
(\ref{DP1}-\ref{DP9}) can as well describe symmetries of a noncommutative spacetime, 
determined by some deformed Heisenberg algebra. Currently it is not yet known how the 
Hopf algebraic structure of symmetries can be extracted from the loop quantization in 3+1 
dimensions, apart from assuming a compatible hypothesis \cite{Amelino-Camelia:2016gfx} (although it has 
recently been done in 2+1 dimensions \cite{Cianfrani:2016ss}). Therefore (\ref{UH}) should be 
understood as the simplest possible Ansatz for the Heisenberg algebra consistent with 
(\ref{DP1}-\ref{DP9}), which does not diverge too far from usual physics. 

\section{Invariant energy scale} \label{IES}

The mass Casimir element (\ref{CasimirLorentz}) is invariant under rotations, since it is constructed 
from the deformation factor (\ref{OmegaLorentz}). Not surprisingly, the set of its symmetries is actually larger, since (\ref{CasimirLorentz}) is also preserved by an appropriate deformed version of the 
Lorentz transformations, which we will now introduce. 

For simplicity let us restrict to boosts with a velocity $v$ 
in the $x$ direction. Then the deformed Lorentz transformation of a four-momentum $(P_0,\{P_a\})$ (whose components are identified with the translation generators via (\ref{eq:II.03})) has the following form:
\begin{align}
P'_0 &= Q \gamma(P_0 - v P_1)\,, \label{L1}\\
P'_1 &= Q \gamma(P_1 - v P_0)\,, \label{L2}\\
P'_2 &= Q P_2\,, \label{L3}\\
P'_3 &= Q P_3\,. \label{L4}
\end{align}
The difference with respect to the ordinary transformations is contained in the factor
\begin{equation}
Q = \frac{1}{\sqrt{1 + \frac{\gamma^2 v^2}{\alpha}\left(P_0^2 + P_1^2 - \frac{2}{v} P_0 P_1\right)}}\,,
\end{equation}
where $\gamma = \frac{1}{\sqrt{1 - v^2}}$ is the usual Lorentz factor. 
In the classical limit $|\alpha| \rightarrow \infty$ we correctly obtain $Q \rightarrow 1$. 

A natural consequence in this context is the existence of an energy scale that is invariant under the 
action of transformations (\ref{L1}-\ref{L4}). Indeed, solving the equation $P'_0 = P_0$ for $P_0$ we find that 
it has two (real-valued) solutions $\pm\sqrt{\alpha}$, although only when $\alpha > 0$. In other words, if we take an arbitrary 
four-momentum of the form $(P_0 = \pm\sqrt{\alpha},P_1,P_2,P_3)$ (with $\alpha > 0$), then acting with a 
deformed boost (\ref{L1}-\ref{L4}) we obtain the vector $(P'_0,P'_1,P'_2,P'_3) = 
(\pm\sqrt{\alpha},Q \gamma(P_1 \mp \sqrt{\alpha}\, v),Q P_2,Q P_3)$, which shows that the energy component 
$P_0$ is conserved. Therefore $\pm\sqrt{\alpha}$ are the invariant energy scales that we were looking for. 
Meanwhile, in the case of $\alpha < 0$ such an invariant is purely imaginary and therefore we do not 
consider it as physical. 

Let us note that $\pm\sqrt{\alpha}$, $\alpha > 0$ are distinguished values of energy within our model for other reasons as well. Namely, both the deformation factor 
$\tilde{\Omega}$ and mass Casimir $\mathcal{C}_1$ become divergent at $|{\bf P}| = \sqrt{\alpha}$. 
Moreover, $\tilde{\Omega}$ changes sign when crossing either $|{\bf P}| = \sqrt{\alpha}$ 
or $|P_0| = \sqrt{\alpha}$. On the other hand, for 
$\alpha < 0$ we observe that $\tilde{\Omega}$ is positive definite, there is no divergence of $\mathcal{C}_1$ and hence no distinguished energy scale. It is also worth to mention that some features of the invariant energy scale $\pm\sqrt{\alpha}$ are similar to the invariant scale characterizing 
one of the models of doubly special relativity \cite{Magueijo:2001cr}. 

Furthermore, it can be expected that the regions in four-momentum space determined by the energy values 
$\pm\sqrt{\alpha}$ -- with $P_0 \in (-\infty,-\sqrt{\alpha})$, $P_0 \in (-\sqrt{\alpha},\sqrt{\alpha})$ or $P_0 \in (\sqrt{\alpha},\infty)$ -- are physically disconnected. More precisely, we want to check whether it is possible to boost a momentum from one region to another. To this end 
let us consider an arbitrary vector of the form $(P_0 = \varepsilon \sqrt{\alpha},P_1,P_2,P_3)$, 
where $\varepsilon \in (-1,1)$, so that $P_0 \in (-\sqrt{\alpha},\sqrt{\alpha})$. A deformed boost with the velocity $v$ transforms the 
energy component into
\begin{equation}
P_0' = \frac{\sqrt{\alpha} \left(\varepsilon \sqrt{\alpha} - v P_1\right)}{\sqrt{\left(\varepsilon \sqrt{\alpha} - v P_1\right)^2 + (1 - v^2) (1 - \varepsilon^2) \alpha}}\,,
\end{equation}
where naturally always $(1 - v^2) (1 - \varepsilon^2) \alpha > 0$ and hence $-\sqrt{\alpha} < P_0' < \sqrt{\alpha}$. Therefore energy remains in the range $(-\sqrt{\alpha},\sqrt{\alpha})$, unless we take superluminal velocities. Choosing $\varepsilon \in (\pm 1,\pm\infty)$ one can reach 
the analogous conclusions for momenta with $P_0$ lying above $\sqrt{\alpha}$ or below $-\sqrt{\alpha}$. These three 
regions can be, therefore, described as physically separated momentum subspaces.

Lastly, let us briefly explore the issue of allowed velocities of particles characterized by the 
considered deformed symmetries. To this end we may consider the mass Casimir (\ref{CasimirLorentz}), 
which gives us the following dispersion relation for particles with mass $m$:
\begin{equation}
P_0^2 = m^2 + {\bf P}^2 \left(1 - \frac{m^2}{\alpha}\right),   
\end{equation}
leading to the following relation between the phase and group velocities:
\begin{equation}
v_{\text{gr}} v_{\text{ph}} = 1 - \frac{m^2}{\alpha}\,. 
\end{equation}
This allows us to express the group velocity as 
\begin{equation}
v_{\text{gr}} = \frac{1 - \frac{m^2}{\alpha}}{\sqrt{1 - \frac{m^2}{\alpha} + \frac{m^2}{{\bf P}^2}}}\,. 
\end{equation}

The maximal allowed value of the group velocity is obtained for  ${\bf P}^2 \rightarrow \infty$ 
and amounts to
\begin{equation}
v_{\text{gr}}^{\text{max}} = \sqrt{1 - \frac{m^2}{\alpha}}\,. 
\end{equation}
Depending on the sign of $\alpha$, $v_{\text{gr}}^{\text{max}}$ can be smaller or greater than 
the speed of light in vacuum. However, for typical masses of particles the difference is expected 
to be very small. Even for the inflaton field, for which $m \sim 10^{-6} m_{\text{Pl}}$, assuming 
$|\alpha| \sim m^2_{\text{Pl}}$ one obtains the correction of the order $\frac{m^2}{|\alpha|} \sim 10^{-12}$. 
The effect would be, therefore, extremely difficult to observe. Moreover, for photons the dispersion 
relation takes the standard form and $v_{\text{gr}}^{\text{max}} = 1$. Anyway, superluminal velocities 
are excluded in the case of $\alpha > 0$.
  
\section{Invariant measure} \label{InvMeas}

When four-momentum space is endowed with a given algebra of symmetries, then the latter determines the form 
of the infinitesimal invariant volume element, which plays the role of a measure on 
this momentum space. For the Poincar\'{e} algebra the invariant 
momentum space measure is simply $d^4P$. On the other hand, since the symmetry algebra (\ref{DP1}-\ref{DP9}) and the corresponding 
transformations (\ref{L1}-\ref{L4}) are a deformed counterpart of the Poincar\'{e} case, one can reasonably expect that the invariant 
measure on momentum space with the symmetries (\ref{DP1}-\ref{DP9}) is an appropriate modification of $d^4P$. 

In order to explore this issue we calculate the Jacobian determinant of 
a momentum transformation (\ref{L1}-\ref{L4})
\begin{equation}
\det\left(\frac{\partial P'_\mu}{\partial P_\nu}\right) =
\left| \begin{array}{cccc} 
\frac{\partial P'_0}{\partial P_0} & \frac{\partial P'_0}{\partial P_1} & \frac{\partial P'_0}{\partial P_2} & \frac{\partial P'_0}{\partial P_3} \\
\frac{\partial P'_1}{\partial P_0} & \frac{\partial P'_1}{\partial P_1} & \frac{\partial P'_1}{\partial P_2} & \frac{\partial P'_1}{\partial P_3} \\
\frac{\partial P'_2}{\partial P_0} & \frac{\partial P'_2}{\partial P_1} & \frac{\partial P'_2}{\partial P_2} & \frac{\partial P'_2}{\partial P_3} \\
\frac{\partial P'_3}{\partial P_0} & \frac{\partial P'_3}{\partial P_1} & \frac{\partial P'_3}{\partial P_2} & \frac{\partial P'_3}{\partial P_3}
\end{array} \right| = Q^6\,.
\end{equation}
The above result leads to the following relation between the measure in the initial coordinates and the one in the boosted coordinates:
\begin{equation}
d^4P' = Q^6 d^4P\,, \label{d4Ptrans}
\end{equation}
where $P = (P_0,\{P_a\})$ and $P' = (P'_0,\{P'_a\})$. 

Let us now try to find such a function $f(P)$ that the condition
\begin{equation}
f(P') d^4P' = f(P) d^4P\,,
\end{equation}
is satisfied. 
From (\ref{d4Ptrans}) we infer that under a deformed boost (\ref{L1}-\ref{L4}) the function $f(P)$ has to transform as
\begin{equation}
f(P) = Q^6 f(P')\,.
\end{equation}
Furthermore, the correspondence with the classical case requires that in the limit $|\alpha| \rightarrow \infty$ we have 
$f(P) \rightarrow 1$. Then the form of $f(P)$ can be deduced by observing that there exists the equality
\begin{equation}
Q^2 \left(1 - \frac{{\bf P}^2}{\alpha}\right) = 1 - \frac{{\bf P'}^2}{\alpha}\,.
\end{equation}
Combining all above relations we find the invariant momentum space measure
\begin{equation}
d\mu \equiv \frac{f(P)d^4P}{(2\pi)^4} = \frac{1}{\left(1 - \frac{{\bf P}^2}{\alpha}\right)^3} \frac{d^4P}{(2\pi)^4}\,.
\label{mommeas} 	
\end{equation}
This result will be crucial for calculating the average return probability in Sec.~\ref{rwsd}. 

It is worth stressing that the measure (\ref{mommeas}) differs from the one that could be naively expected 
from the mass Casimir (\ref{CasimirLorentz}). Namely, writing (\ref{CasimirLorentz}) as
\begin{equation}
\mathcal{C}_1 = g^{\mu\nu}(P) P_{\mu} P_{\nu}\,, \quad g^{\mu\nu}(P) \equiv \frac{\eta^{\mu\nu}}{1 - \alpha^{-1} {\bf P}^2}\,,
\end{equation}
one naturally deduces that the invariant measure should have the form 
\begin{equation}
\sqrt{|\det(g_{\mu\nu}(P))|}\, d^4P = \left(1 - \frac{{\bf P}^2}{\alpha}\right)^2 d^4P\,.
\end{equation}
However, the measure obtained in such a heuristic way is explicitly breaking the 
invariance with respect to the deformed boosts (\ref{L1}-\ref{L4}).

\section{Euclidean domain}

So far we have focused on the Lorentzian model, with the signature $s = 1$. However, to introduce a diffusion process on spacetime we actually need to 
consider the Euclidean version of our deformed symmetry algebra. 
One of the issues in this context is that the Laplace operator defined on spacetime (as well as in the momentum space representation) 
has to be elliptic. For the model considered in this paper the situation is 
complicated by the fact that the sign of $\tilde{\Omega}$ changes with momentum. Therefore, we expect 
the Laplace operator to be of a mixed type (either elliptic 
or hyperbolic, depending on the energy range). In order to be able 
to probe spacetime by diffusion from large scales up to small scales we restrict here to the situation where spacetime is Euclidean 
at large scales. Nevertheless, one has to keep in mind that when $\alpha > 0$ the sign of the Euclidean version of $\tilde{\Omega}$ changes and spacetime becomes 
Lorentzian for 3-momenta exceeding $\sqrt{\alpha}$, which is making the diffusion ill defined for such a regime of momenta. The alternative possibility would be to consider 
the Lorentzian case discussed in Sec.~\ref{DefPoinc}, again with the sign changing 
$\tilde{\Omega}$. Such a case is, however, also ill defined (for $\alpha > 0$) since 
it would contain the regime where the measure is negative, leading to a substantial difficulty 
in interpreting the trace of heat kernel as the average return probability (see Sec.~\ref{rwsd}). 
The latter could be remedied \cite{Calcagni:2013vsa} by an appropriate modification of the 
diffusion equation but here we follow the conservative approach. 

A transition from the Lorentzian to Euclidean domain can be performed either
by an analytic continuation, called the Wick rotation, or by introducing a priori 
the Euclidean counterpart of the deformed Poincar\'{e} algebra. Since the   
analytic continuation may turn out to be tricky, as it is the case for the 
$\kappa$-Poincar\'{e} algebra (where the deformation parameter $\kappa$ has to be analytically continued as well, although this is so due to the coalgebra \cite{Lukierski:1994qt}), 
both of the possibilities are discussed below. 

The Wick rotation that we consider here is the analytic continuation 
$P_0 \longrightarrow -i P_0$, $K_a \longrightarrow -i K_a$. Applying it to the deformed brackets (\ref{DP1}-\ref{DP9}) we find that the only formulae that change are those depending on the effective signature $s_{\text{eff}} = s \tilde{\Omega}$, which become
\begin{align}
\left\{K_a,K_b \right\} &= s_{\text{eff}} \epsilon_{abc} J^c\,, \\
\left\{K_a,P_0 \right\} &= -s_{\text{eff}} P_a\,.
\end{align}
By redefining the signature $s$ to the correct Euclidean value $s = -1$ they can be restored to their previous form (\ref{DP3}), (\ref{DP7}) but with the Wick-rotated $\tilde{\Omega}$. Namely, applying $P_0 \longrightarrow -i P_0$ to the Lorentzian deformation factor (\ref{OmegaLorentz}) 
we obtain its Euclidean counterpart
\begin{equation}
\tilde{\Omega}^E = -\frac{P^2_0 + \alpha}{{\bf P}^2 - \alpha} \label{OmegaEuclid1}
\end{equation}
and therefore the effective signature turns into $s_{\text{eff}} = -\tilde{\Omega}^E$. Then the brackets (\ref{DP1}-\ref{DP9}) define the deformed Euclidean algebra, while in the low energy limit $s\tilde{\Omega}^E(0,0) \rightarrow -1$ and hence the classical Euclidean algebra is recovered. 

Furthermore, Wick-rotating the Lorentzian Casimir element (\ref{CasimirLorentz}) we obtain
\begin{equation}
\mathcal{C}^E_1 = \frac{P_0^2 + {\bf P}^2}{1 - \alpha^{-1}{\bf P}^2}\,. 
\label{CasimirEuclid1}
\end{equation}
One can observe that for $\alpha < 0$ the expression (\ref{CasimirEuclid1}) is a positive definite function, while 
for $\alpha > 0$ it is positive below the invariant energy scale $|{\bf P}| < \sqrt{\alpha}$, negative
above this scale $|{\bf P}| > \sqrt{\alpha}$ and divergent at $|{\bf P}| = \sqrt{\alpha}$. The negativity of $\mathcal{C}^E_1$ can indicate that we are entering the hyperbolic regime of the Laplace operator. 
However, this is so only if the relation between the Laplace operator 
and the mass Casimir is linear, while in general this may not be the case. 

In order to explore the other possible definition of the Euclidean counterpart of 
our model we first assume that $s = -1$. In this case the classical limit gives us the condition 
$s \tilde{\Omega}^E(0,0) = -1$, equivalent to $\tilde{\Omega}^E(0,0) = 1$. The latter 
is the same as in the Lorentzian case for the function $\tilde{\Omega}$. 
Consequently, the solution for $\tilde{\Omega}^E$ is analogous to (\ref{OmegaSolution}) 
and reads 
\begin{equation}
\tilde{\Omega}^E(P_0,{\bf P}^2) = \frac{P^2_0/c_1 - 1}{{\bf P}^2/c_2 - 1}\,. 
\label{OmegaEucSol} 
\end{equation}
The agreement with the expression (\ref{OmegaEuclid1}) can be achieved by setting 
$c_2 = -c_1 = \alpha$, for which (\ref{OmegaEucSol}) reduces to 
\begin{equation}
\tilde{\Omega}^E = - \frac{P^2_0 + \alpha}{{\bf P}^2 - \alpha}\,. \label{OmegaEuclid2}
\end{equation}
Then the mass Casimir is again given by a 
superposition of $\tilde{\Omega}^E$ and the unit element, namely
\begin{equation}
\mathcal{C}^{E}_1 = a_1\tilde{\Omega}^{E} + a_2 \mathbb{I} =
\frac{a_1 P_0^2 + a_2 {\bf P}^2 + \alpha (a_1 - a_2)}{{\bf P}^2 - \alpha}\,.
\end{equation} 
The conditions that guarantee the proper classical limit $\mathcal{C}^E_1 = P^2_0 + {\bf P}^2$ 
are $a_2 = -a_1$ and $a_1 = \alpha$, leading to   
\begin{equation}
\mathcal{C}^E_1 = 
\frac{P^2_0 + {\bf P}^2}{1 - \alpha^{-1} {\bf P}^2}\,, 
\label{CasimirEuclid2} 
\end{equation} 
which is identical to (\ref{CasimirEuclid1}), as it should.

\section{Diffusion and the spectral dimension} \label{rwsd}

We have already collected all necessary ingredients to address the main subject of this 
paper -- a diffusion process on (Wick-rotated) spacetime with the deformed symmetries (\ref{DP1}-\ref{DP9}). Namely, on a Riemannian manifold of $d$ topological 
dimensions and with the metric $g$ a diffusion (or random walk) is described by the heat equation
\begin{equation}
\frac{\partial}{\partial\sigma} K(x,x_0;\sigma) = \Delta_x K(x,x_0;\sigma)\,, 
\label{HeatKernel}
\end{equation}
where $\sigma$ denotes the auxiliary time (which plays the role of a scale parameter), $\Delta_x$ is the Laplacian and we assume the initial condition
\begin{equation}
K(x,x_0; \sigma = 0) = \frac{\delta^{(d)}(x - x_0)}{\sqrt{|\det g(x)|}}\,.
\end{equation}
In general, the Laplacian may differ from the usual one 
$\Delta = g^{\mu\nu} \partial_\mu \partial_\nu$, $\mu,\nu = 1,\ldots,d$. For $\mathbb{R}^4$ with 
the standard Euclidean metric the solution to (\ref{HeatKernel}), also called the heat kernel,
can be written as
\begin{equation}
K(x,x_0;\sigma) = \int \frac{d^4P}{(2\pi)^4} e^{i P_\mu (x - x_0)^\mu} e^{\sigma \Delta_P}\,, \label{HKSol}
\end{equation}
where $\Delta_P$ is the Laplacian represented on space of momenta $P$. As one can see, the expression (\ref{HKSol}) is obtained using the Fourier transform.  

For the model discussed in Sec.~\ref{DefPoinc} we have shown that the deformed 
Poincar\'{e} algebra (\ref{DP1}-\ref{DP9}) can be consistently
complemented by the undeformed Heisenberg algebra (\ref{UH}) (although it is not the unique choice). Then phase space has the ordinary structure and it seems that the usual notion of the Fourier transform should be preserved. However, in the case of models characterized by non-trivial phase space the Fourier transform can be substantially different from the standard one \cite{Guedes:2013qs}. In particular, it may lead to a modification of the measure on momentum space, as it happens for the $\kappa$-Poincar\'{e} algebra \cite{Freidel:2008fy}. By analogy, the deformed measure (\ref{mommeas}) from Sec.~\ref{InvMeas} can perhaps be coming from the corresponding modification of the Fourier transform, which arises due to the deformation of symmetries (\ref{DP1}-\ref{DP9}), irrespective of the form of the Heisenberg algebra. 

In order to correctly define the counterpart of the heat kernel (\ref{HKSol}) for the considered model we have to make two changes that we have already suggested. Namely, use the appropriate Laplace operator $\Delta_P$ and the 
invariant measure on momentum space $d\mu(P)$. The invariance is necessary to guarantee the independence of the results from a 
particular reference frame, which has to be fulfilled at the level of the average 
return probability
\begin{align}
P(\sigma) &:= {\rm tr} K(\sigma) = \frac{\int \sqrt{|\det g|}\, d^4x K(x,x;\sigma)}{\int \sqrt{|\det g|}\, d^4x} 
\nonumber\\
&= \int d\mu(P)\, e^{\sigma \Delta_P}\,, \label{retprob}
\end{align}
obtained by the space averaging of the return probability $K(x,x;\sigma)$. Both 
$\Delta_P$ and $d\mu(P)$ should be invariant quantities, leading to the invariance 
of the whole expression. Then the spectral dimension is defined as
\begin{equation}
d_S(\sigma) := -2 \frac{\partial \log P(\sigma)}{\partial \log \sigma}
\label{SpectralDim}
\end{equation}
and is a function of the scale parameter $\sigma$. Nevertheless, in our case $g$ is the Euclidean metric and hence $P(\sigma) = K(x,x;\sigma)$. 

The invariance of the Laplace operator $\Delta_P$ under rotations and deformed boosts is 
guaranteed by the fact that it can be expressed in terms of the Euclidean mass 
Casimir as
\begin{equation}
\Delta_P = -\mathcal{C}^E_1 + \sum_{n = 2} c_n \alpha \left(\frac{\mathcal{C}^E_1}{\alpha}\right)^n.
\end{equation}
The last term has been included to show that, in general, higher order powers of the Casimir may contribute 
to the expression for $\Delta_P$. However, here we restrict to 
the simplest possibility, setting $\forall_{n \geq 2} c_n = 0$, simply due 
to the fact that we do not know what the extra $c_n$ parameters could be.  

Taking $\Delta_P = -\mathcal{C}^E_1$ given by 
(\ref{CasimirEuclid2}) and the invariant measure (\ref{mommeas}) we can finally write the following average return probability:
\begin{equation}
P(\sigma) = \frac{4\pi}{(2\pi)^4} \int_{I_p} 
\frac{dp\, p^2}{\left(1 - \frac{p^2}{\alpha}\right)^3} \int_{I_E} dE e^{-\sigma \frac{E^2 + p^2}{1 - p^2/\alpha}}\,.
\label{Psigma}
\end{equation}
Here we introduce the notation $E \equiv P_0$ and $p \equiv |{\bf P}|$, while $I_E$ and $I_p$ denote the 
corresponding ranges of integration. The above expression can be given the probabilistic interpretation when $P(\sigma) > 0$. It also has the correct classical limit
\begin{equation}
\lim_{|\alpha| \rightarrow \infty} P(\sigma) = \frac{1}{16\pi^2} \frac{1}{\sigma^2}\,, \label{ClassLim}
\end{equation}
for which one obtains $d_S(\sigma) = 4$.

In the remaining part of this Section we will calculate the explicit form of $P(\sigma)$ with either sign 
of the parameter $\alpha$ and the appropriate intervals $I_E$ and $I_p$. From each of the results we will extract the corresponding 
spectral dimension (\ref{SpectralDim}).
 
\subsection{The case $\alpha > 0$, $I_p = [0,\sqrt{\alpha}]$, $I_E = \mathbb{R}$}

At the beginning let us consider two cases with positive $\alpha$. As we discussed 
in Sec.~\ref{IES}, then 
$\pm\sqrt{\alpha}$ are the invariant energy scales. Furthermore, at $p = \sqrt{\alpha}$ (note that $p > 0$) the Laplace operator $\Delta_P$ 
changes its type from elliptic to hyperbolic. For this reason the integration 
range of $p$ has to be restricted to $I_p = [0,\sqrt{\alpha}]$. On the other hand, crossing of the value 
$E = \sqrt{\alpha}$ does not 
change sign of the mass Casimir (\ref{CasimirEuclid1}) and therefore we keep here the full energy range $I_E = \mathbb{R}$. A case with the bounded $I_E$ will be discussed in the next Subsection. Applying the above 
integration intervals to (\ref{Psigma}) we obtain the average return probability
\begin{align}
P(\sigma) &= \frac{4\pi}{(2\pi)^4} \sqrt{\frac{\pi}{\sigma}}  \int_{0}^{\sqrt{\alpha}}
\frac{dp\ p^2 e^{-\sigma \frac{p^2}{1 - p^2/\alpha}}}{\left(1 - \frac{p^2}{\alpha}\right)^{5/2}} \nonumber \\
&= \frac{4\pi}{(2\pi)^4} \sqrt{\frac{\pi}{\sigma}} \int_0^{\infty} dq\, q^2 e^{-\sigma q^2} \nonumber \\
&= \frac{1}{16\pi^2} \frac{1}{\sigma^2}\,, \label{eq:VI.01}
\end{align}
where the change of variables to $q := \frac{p}{\sqrt{1 - p^2/\alpha}}$ has converted the seemingly 
complicated formula into the case (\ref{ClassLim}) of Euclidean space with the ordinary Laplacian. Substituting this result into (\ref{SpectralDim}) we calculate the spectral dimension
\begin{equation}
d_S(\sigma) = 4\,.
\end{equation}  
Therefore, for $\alpha < 0$ and the integration over momentum space restricted to $I_p = [0,\sqrt{\alpha}]$, 
we measure the same dimension of (Wick-rotated) spacetime as for the four-dimensional Minkowski spacetime. However, this not necessarily means that the two spaces are isomorphic to each other. 

\subsection{The case $\alpha > 0$, $I_p = [0,\sqrt{\alpha}]$, $I_E = [-\sqrt{\alpha},\sqrt{\alpha}]$}

\begin{figure}[ht!]
\centering
\includegraphics[width=7cm,angle=0]{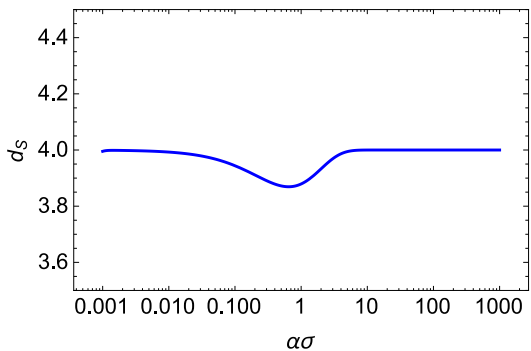}
\caption{Spectral dimension as a function of the scale parameter for the 
case with $\alpha > 0$ and $I_E = [-\sqrt{\alpha},\sqrt{\alpha}]$.} 
\label{SpectrRunAlph+}
\end{figure}

Another possibility is that, in contrast to the previous Subsection, energy is restricted to the range between the invariant 
scales, i.e.\! $I_E = [-\sqrt{\alpha},\sqrt{\alpha}]$. Then, performing the calculations analogously to (\ref{eq:VI.01}), we find that (\ref{Psigma}) has the form
\begin{align}
P(\sigma) = \frac{4\pi}{(2\pi)^4} \sqrt{\frac{\pi}{\sigma}} \int_0^\infty dq\, q^2 e^{-\sigma q^2} 
{\rm erf}\left(\sqrt{\sigma \left(q^2 + \alpha\right)}\right)\,, \label{eq:VI.02}
\end{align}
with the following asymptotic behaviors:
\begin{equation}
P(\sigma) \sim \sigma^{-2} \quad \text{for} \quad \sigma \rightarrow \infty\,, 
\end{equation}
and 
\begin{equation}
P(\sigma) \sim \sigma^{-2} \quad \text{for} \quad \sigma \rightarrow 0\,. 
\end{equation}
The spectral dimension corresponding to (\ref{eq:VI.02}) can only be obtained numerically and its running is shown in Fig.~\ref{SpectrRunAlph+}.
Curiously, the restriction of energy to $[-\sqrt{\alpha},\sqrt{\alpha}]$ leads here to a slight 
deviation from $d_S = 4$ at intermediate diffusion times $\sigma \sim 1/\alpha$, characterized by the invariant energy scale. In principle we could also consider the cases with $I_E = [-\infty,-\sqrt{\alpha}]$ or $I_E = [\sqrt{\alpha},\infty]$. However, this would not allow us to probe the large scale limit of spacetime, which is associated with 
low energy contributions to the Laplace operator. 

\subsection{The case $\alpha < 0$, $I_p = \mathbb{R}$, $I_E = \mathbb{R}$}

Let us now turn to the case of $\alpha < 0$. For the Lorentzian signature $s = 1$ the sign 
of $\tilde{\Omega}$ remains constant irrespective of the values of $E$ and $p$, 
and hence the signature change does not occur. In turn, the Euclidean $\tilde{\Omega}^E$ 
changes its sign at $P^2_0 = -\alpha$.  Nevertheless, the Laplace operator $\Delta_P$ remains 
elliptic in the whole range of $E$ and $p$ and we have no reason to restrict the intervals $I_E$ 
and $I_p$. Similarly, there is no divergence in either $\tilde{\Omega}^E$ or $\Delta_P$ 
and, which will turn out to be essential, the value of $\tilde{\Omega}^E$ tends to 
zero in the ${\bf P} \rightarrow \infty$ limit. 

\begin{figure}[ht!]
\centering
\includegraphics[width=7cm,angle=0]{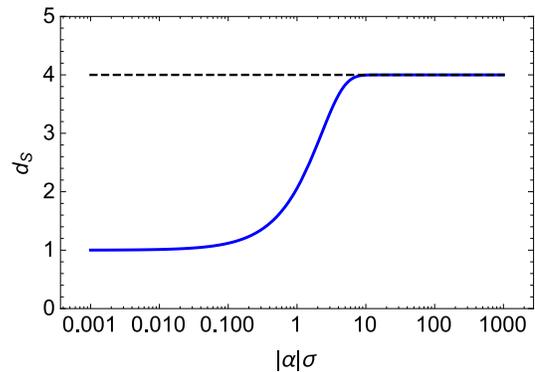}
\caption{Spectral dimension as a function of the scale parameter for the 
case with $\alpha < 0$ with $I_E =\mathbb{R}$.} 
\label{SpectrRunAlph-}
\end{figure}

Integrating (\ref{Psigma}) over $E$ and making the change of variables to 
$u := \frac{p}{\sqrt{1 + p^2/|\alpha|}}$ we obtain
\begin{align}
P(\sigma) &= \frac{4\pi}{(2\pi)^4} \sqrt{\frac{\pi}{\sigma}} \int_0^\infty
\frac{dp\, p^2 e^{-\sigma \frac{p^2}{1 + p^2/|\alpha|}}}{\left(1 + \frac{p^2}{|\alpha|}\right)^{5/2}} \nonumber\\
&= \frac{4\pi}{(2\pi)^4} \sqrt{\frac{\pi}{\sigma}} \int_0^{\sqrt{|\alpha|}} du\, u^2 e^{-\sigma u^2} \nonumber\\
&= \frac{{\rm erf}(\sqrt{|\alpha| \sigma})}{16\pi^2 \sigma^2} - \frac{e^{-|\alpha| \sigma} \sqrt{|\alpha|}}{8\pi^{5/2} \sigma^{3/2}}\,, \label{eq:VI:03}
\end{align}
which has the asymptotic behaviors:
\begin{equation}
P(\sigma) \sim \sigma^{-2} \quad \text{for} \quad \sigma \rightarrow \infty\,,
\end{equation}
and
\begin{equation}
P(\sigma) \sim \sigma^{-1/2} \quad \text{for} \quad \sigma \rightarrow 0\,.
\end{equation}
The analytic expression for the running spectral dimension calculated from (\ref{eq:VI:03}) is
\begin{equation}
d_S(\sigma) = 4 - \frac{4 (|\alpha| \sigma)^{3/2}}{\sqrt{\pi}\, e^{|\alpha| \sigma} {\rm erf}(\sqrt{|\alpha| \sigma}) - 2 \sqrt{|\alpha| \sigma}}
\end{equation}
and the corresponding plot is presented in Fig.~\ref{SpectrRunAlph-}.

\begin{figure}[ht!]
\centering
\includegraphics[width=7cm,angle=0]{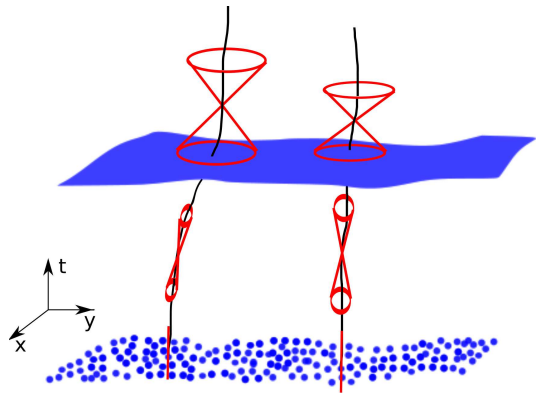}
\caption{Schematic representation of the collapse of lightcones into a congruence of null worldlines while 
approaching the ultralocal (silent) state.} 
\label{AsSil}
\end{figure}

We observe here the dimensional reduction from the large scale value $d_S = 4$ to $d_S = 1$ at small scales. The latter 
ultraviolet value can have the following interpretation. At high energies $|{\bf P}| \rightarrow \infty$ 
the deformation factor $\tilde{\Omega}$ tends to zero, corresponding to the so-called ultralocal \cite{Isham:1975ur} or Carrollian 
limit \cite{Carroll:1965,Bacry:1968ps} of spacetime and its relativistic symmetries. In the Carrollian limit, in which the speed of light is taken to zero, spacelike separated points become effectively 
decoupled, due to the collapse of their lightcones into null worldlines. Then spacetime becomes a congruence 
of such (one-dimensional) worldlines, as it is schematically depicted in Fig.~\ref{AsSil}. Since increasing energy is equivalent to probing smaller scales of spacetime, we indeed have the 
dimensional reduction to $d_S = 1$. Such a value is consistent with the analysis \cite{Mielczarek:2012tn} 
of loop quantum cosmology in the case characterized by the hypersurface deformation algebra 
(\ref{DHDA1}-\ref{DHDA3}). The significance of these results is that ultralocality is one of the features of 
the Belinsky-Khalatnikov-Lifshitz (BKL) conjecture \cite{Belinsky:1970ew} or asymptotic silence scenario. 

On the other hand, in \cite{Carlip:2016dy} 
it has been argued that asymptotic silence is characterized by the dimensional reduction to 2, 
which results from an elongation of the anisotropic cosmological model in one particular direction 
during each of the Kasner epochs. Then spacetime effectively has one spatial and one temporal direction. 
However, such a viewpoint is in contradiction with ultralocality, which is achieved first in the BKL scenario. 
Namely, in this scenario universe first decouples into non-interacting points of space, each described by the minisuperspace 
homogeneous cosmological model. Then the chaotic dynamics is 
acting at the level of the points. Treating an elongation in the internal space of a point as an elongation in the three-dimensional 
space violates the initial suppression of the spatial dependence of fields. Therefore, the interpretation 
in which the BKL conjecture leads to the dimensional reduction to 1 seems to be better justified. 

It should be stressed that the results in this Subsection (as well as in the previous ones) have been obtained 
in the Euclidean domain of the model, which can be treated either as an independent case or as the Wick-rotated version of the Lorentzian model. The second possibility is supported by the observed consistency with the Lorentzian picture 
of the collapse of lightcones. Therefore, we 
conclude that the dimensional reduction of our model does not fundamentally depend 
on the spacetime metric signature. This is also in agreement with the fact that, irrespective of the signature $s$, in the limit $\tilde{\Omega} \rightarrow 0$ (or $\tilde{\Omega}^E \rightarrow 0$) the effective 
signature $s_{\text{eff}} = s \tilde{\Omega}$ (or $s_{\text{eff}} = s \tilde{\Omega}^E $) tends 
to zero and the deformed algebra (\ref{DP1}-\ref{DP9}) becomes the (standard) Carroll algebra. 

\subsection{The case $\alpha < 0$, $I_p = \mathbb{R}$, $I_E = [-\sqrt{\alpha}, \sqrt{\alpha}]$}

While the Laplace operator for $\alpha < 0$ is elliptic independently of the value of $P_0$, 
in order to make our discussion complete we will also consider here the restriction to 
$I_E = [-\sqrt{\alpha}, \sqrt{\alpha}]$. The motivation for taking into account such a range of 
variability of $P_0$ is the changing sign of $s_{\text{eff}}$. Namely, for the Euclidean case, 
with the deformation factor (\ref{OmegaEuclid1}), we find that
\begin{equation}
s_{\text{eff}} = -\tilde{\Omega}^E = \frac{P^2_0 + \alpha}{{\bf P}^2 - \alpha}. 
\end{equation}
For $\alpha < 0$ the denominator remains positive definite and the effective signature $s_{\text{eff}}$ is 
negative (Euclidean) for $P^2_0 < |\alpha|$, while for $P^2_0 = |\alpha|$ we obtain $s_{\text{eff}} = 0$ and 
for $P^2_0 > |\alpha|$ the signature $s_{\text{eff}}$ becomes positive (Lorentzian). Therefore, while entering 
the $P^2_0 > |\alpha|$ region does not affect the elliptic character of the Laplace operator, the effective signature 
becomes Lorentzian. Below we make an analysis of the diffusion process assuming that the $P^2_0 > |\alpha|$ 
regime is excluded from the physical phase space. In other words, a UV cut-off is introduced at the energy scale 
$P_0 = \sqrt{|\alpha|}$.  

\begin{figure}[ht!]
\centering
\includegraphics[width=7cm,angle=0]{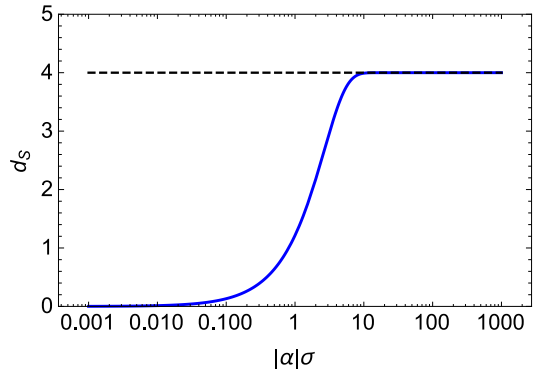}
\caption{Spectral dimension as a function of the scale parameter for the 
case with $\alpha < 0$  with $I_E = [-\sqrt{\alpha}, \sqrt{\alpha}]$.} 
\label{SpectrRunAlphCaseD}
\end{figure}

In this case the average return probability can be written as
\begin{equation}
P(\sigma) = \frac{|\alpha|^2 e^{-\frac{\tau}{2}}}{16 \pi^2 \tau} 
\int_{-1}^{1} \frac{dx\, e^{-\tau x^2 /2}}{1 - x^2}\, {\rm I}_1\left(\frac{\tau}{2} (1 - x^2)\right),
\label{PcaseD}
\end{equation}
where $\tau := \sigma |\alpha|$ and ${\rm I}_1(x)$ denotes a Bessel function. 
The expression (\ref{PcaseD}) has the following asymptotic behaviors:
\begin{equation}
P(\sigma) \sim \sigma^{-2} \quad \text{for} \quad \sigma \rightarrow \infty\,,
\end{equation}
and
\begin{equation}
P(\sigma) \sim {\rm const} \quad \text{for} \quad \sigma \rightarrow 0\,.
\end{equation}
The UV behavior suggests that the spectral dimension reduces to $d_S = 0$ in 
this limit. This expectation is supported by calculating $d_S$ as 
a function of $\sigma$ with the use of the definition (\ref{SpectralDim}). 
The resulting $d_S(\sigma)$ dependence is plotted in Fig.~\ref{SpectrRunAlphCaseD}.

How to understand such a result? In the case with $I_E = \mathbb{R}$ the spectral dimension 
in the UV limit reduces to $d_S = 1$, which, as we already discussed, can be associated 
with the time direction of spacetime. Introducing the cut-off $\sqrt{|\alpha|}$ in the energy domain (which 
is the case discussed in the present Subsection) prevents the probing of the time direction at 
time scales smaller than $1/\sqrt{|\alpha|}$. In other words, when the energy cut-off is 
introduced, time effectively becomes discrete and undefined at the scales smaller than 
$1/\sqrt{|\alpha|}$. Therefore, it is natural to expect that the time dimension (which remains physically available 
in the ultralocal case) disappears at sufficiently short time scales if the energy cut-off is introduced.

\section{Summary}

The analysis presented in this paper addressed the issue of probing a loop-quantized spacetime configuration 
by the diffusion process. The considered model was given by Minkowski spacetime but with symmetries described by 
a deformed Poincar\'{e} algebra, whose form is motivated by the predictions of the effective regime of loop 
quantum gravity. While such an algebra is still unknown in general, we proposed a specific case that has the 
structure recovered from the loop-deformed hypersurface deformation algebra. Namely, using reasonable 
assumptions, like the conservation of the Jacobi identities by the deformed Poincar\'{e} algebra, we have 
determined the particular form of the deformation factor. We also assumed an extension of such a symmetry 
algebra by the standard Heisenberg algebra of phase space variables. 

Subsequently we showed that our symmetry algebra leads to deformed Lorentz transformations, which 
preserve two distinguished energy scales. Moreover, we have derived the form of the invariant measure 
on momentum space. The above ingredients were applied to precisely define (after the Wick rotation) and 
analyze the diffusion process on spacetime endowed with the considered deformed algebra of symmetries. 
In particular, we have found that in the case in which the deformation factor tends to zero in the high energy 
limit, the spectral dimension reduces to $d_S = 1$ at small scales, as it is expected in the asymptotic silence 
scenario. Besides, the presence of the invariant energy scale allows to consider our model as an example of 
the doubly special relativity (DSR), which could be further studied within the latter framework. 

The UV values of the spectral dimension for all the cases considered 
in this article are collected in the table below. 

\begin{table}[ht]
\centering
\begin{tabular}{|c|c|c|c|c|}
\hline
Case & $\alpha$ & $I_p$ & $I_E$  & $d_S(\sigma \rightarrow 0)$ \\ [0.5ex] \hline
A & $>0$ & $[0, \sqrt{\alpha}]$ & $\mathbb{R}$ & 4 \\
B & $>0$ & $[0, \sqrt{\alpha}]$ & $[-\sqrt{\alpha}, \sqrt{\alpha}]$  & 4 \\
C & $<0$ & $\mathbb{R}$ & $\mathbb{R}$ & 1 \\
D & $<0$ & $\mathbb{R}$ & $[-\sqrt{\alpha}, \sqrt{\alpha}]$ & 0 \\
\hline
\end{tabular}
\label{table:dimres}
\end{table}

Finally, let us make a brief comparison of our derivation of the deformed Poincar\'{e} algebra 
with the LQG derivation in the spherically symmetric case in \cite{Bojowald:2012ux,Amelino-Camelia:2016gfx}. 
In this paper we have begun with assuming a general deformation of the bracket (\ref{DHDA3}), parametrized 
by some unknown function $\Omega$ of phase space variables. As it should be stressed, the 
subsequent reduction of the DHDA to the corresponding deformed Poincar\'{e} algebra requires the use 
of a certain averaging procedure. The obtained deformation of the Poincar\'{e} algebra is then parametrized 
by the $s_{\text{eff}}$ function, whose form is inferred from the consistent mathematical assumptions 
rather than some known physical solution for $\Omega$. The family of functions $s_{\text{eff}}$ considered 
in this paper does not cover all possible choices of $\Omega$. Moreover, it is not yet known if the 
discussed form of $s_{\text{eff}}$ can be associated with the cosine form of $\Omega$ that appears in the 
LQG models with holonomy corrections. In contrast, in \cite{Bojowald:2012ux,Amelino-Camelia:2016gfx} 
the cosine form of $\Omega$ is suggested by the presence of the calculated holonomy 
corrections. It is subsequently assumed that the corresponding deformation factor of the Poincar\'{e} 
algebra is of the same cosine form. However, as we have to stress, a straightforward identification 
of the form of HDA deformation function $\Omega$ with the corresponding deformation 
of the Poincar\'{e} algebra is not necessarily correct and the known results in the spherically symmetric 
case need to be confirmed by the further analysis. Due to such differences between both of the discussed  approaches, at the moment we are only able to observe the qualitative but not the quantitative similarities 
between their results.
 
\section*{Acknowledgements.} We thank Anna Pacho\l\ for very useful discussions. 
This work is supported by the Iuventus Plus grant No.~0302/IP3/2015/73 from the 
Polish Ministry of Science and Higher Education. TT was additionally supported by 
the National Science Centre Poland, project 2014/13/B/ST2/04043.  

\section*{Appendix}

The aim of this Appendix is to show explicitly how the classical hypersurface deformation 
algebra reduces to the Poincar\'{e} or Euclidean algebra, depending on the metric signature $s$. 
A more detailed discussion of these calculations can be found e.g.\! in \cite{Thiemann:2008}. 

The hypersurface deformation algebra describes deformations of an arbitrary spatial 
hypersurface $\Sigma_t$, which correspond to local diffeomorphisms. The generators 
of a deformation are the smeared scalar constraint $S[N]$ and the smeared spatial 
diffeomorphisms constraint $D[N^a]$. $S[N]$ generates deformations in the direction 
normal to the hypersurface $\Sigma_t$ (i.e.\! the time direction), while $D[N^a]$ is 
responsible for deformations in the tangential direction. The deformations are 
parametrized by a lapse function $N$ and shift vector $N^a$, which together form 
the deformation vector $u^\mu = N n^\mu + N^\mu$, where $n^\mu$ is a unit vector 
normal to the hypersurface $\Sigma_t$, such that $g_{\mu\nu} n^\mu n^\nu = s$. Here 
$\mu,\nu = 0,1,2,3$ denote spacetime indices and $a,b = 1,2,3$ the spatial ones. The 
local coordinate transformation generated by $u^\mu$ is given by $x^\mu \rightarrow x'^\mu 
= x^\mu + u^\mu$. 

The constraints $S[N]$ and $D[N^a]$ are contributions to the gravitational 
Hamiltonian in the ADM formulation, $H[N,N^a] = S[N] + D[N^a]$. Furthermore, they form the first class algebra:
\begin{align}
\left\{ D[N^a_1],D[N^a_2] \right\} &= D\left[ N_1^b \partial_b N^a_2 - N_2^b \partial_b N^a_1 \right], \label{HDA1} \\
\left\{ S[N],D[N^a] \right\} &= -S\left[ N^b \partial_b N \right], \label{HDA2} \\
\left\{ S[N_1],S[N_2] \right\} &= D\left[ s q^{ab} (N_1 \partial_b N_2 - N_2 \partial_b N_1) \right]. \label{HDA3}
\end{align}
The above algebra is satisfied 
by any theory that is covariant under local diffeomorphisms. Let us note that due to the presence of the spatial metric $q^{ab}$ in the last line the 
brackets (\ref{HDA1}-\ref{HDA3}) do not define a Lie algebra. 

A special class of local diffeomorphisms are linear transformations, which are 
associated with the Poincar\'{e} (or respectively Euclidean) symmetry. The corresponding brackets of the symmetry 
generators can be recovered from the hypersurface deformation algebra (\ref{HDA1}-\ref{HDA3}) 
by restricting the (infinitesimal) deformations to linear functions, such that we have
\begin{align}
N(x) &= \Delta t + v_a x^a\,, \label{lin1}\\
N^a(x) &= \Delta x^a + {R^a}_b x^b\,, \label{lin2}\\
q_{ab} &= \delta_{ab}\,, \label{lin3}
\end{align}
where $\Delta t$ is the parameter of a time translation, $v_a$ parametrize 
boosts, $\Delta x^a$ specify spatial translations and ${R^a}_b$ is a rotation matrix. The 
rotation matrix in terms of infinitesimal angles $\varphi^a$ is given 
by $R^{ab} = \epsilon^{bac} \varphi_c$. 

In the linear case (\ref{lin1}-\ref{lin3}) the constraints $D[N^a]$ and $S[N]$ can be 
expressed as the following combinations of the Poincar\'{e} (or respectively Euclidean) generators:
\begin{align}
D[N^a] &= -\Delta x^a P_a-\varphi^a J_a\,, \\
S[N] &= -\Delta t P_0 - v^a K_a\,, 
\end{align}
where $P_0$ is the generator of time translations, $P_a$ are the generators of spatial 
translations, $J_a$ generators of rotations and $K_a$ generators of boosts. The minus 
signs are necessary to obtain the symmetry generators in the usual convention. Then the left hand side of the bracket (\ref{HDA1}) can be rewritten as
\begin{align}
\left\{ D[N^a_1],D[N^a_2] \right\} &= \left\{ \Delta x_1^a P_a + \varphi_1^a J_a ,\Delta x_1^b P_b + \varphi_1^b J_b \right\} \nonumber\\
&= \Delta x_1^a\Delta x_2^b \left\{ P_a,P_b\right\} \nonumber\\
&+ (\varphi^a_1 \Delta x^b_2 - \varphi^a_2 \Delta x^b_1) \left\{ J_a,P_b \right\} \nonumber\\
&+ \varphi^a_1 \varphi^b_2 \left\{ J_a,J_b \right\}, \label{DDL}
\end{align}
whereas the right hand side simplifies to
\begin{align}
& D\left[ N_1^b \partial_b N^a_2 - N_2^b \partial_b N^a_1 \right] =  \nonumber \\
&= D[(\Delta x_1^b \varphi_{2c} - \Delta x_2^b \varphi_{1c}) \epsilon^{acb} - \epsilon^{abc} (\epsilon_{bde} \varphi_1^d \varphi_2^e)x_c] \nonumber \\
&= (\Delta x_2^b \varphi_{1c} - \Delta x_1^b \varphi_{2c}) \epsilon^{acb} P_a + \epsilon^{bde} \varphi_{1d} \varphi_{2e} J_b\,. 
\label{DDR}
\end{align}
The final expressions in (\ref{DDL}) and (\ref{DDR}) agree with (\ref{HDA1}) if and only if the appropriate terms on both sides (which are multiplied by parameters of the deformations) are equal to each other. 
This condition gives us the first three brackets of the Poncar\'{e} algebra:
\begin{align}
\left\{P_a,P_b\right\} &= 0\,, \label{P1} \\
\left\{J_a,P_b\right\} &= \epsilon_{abc}P^c\,, \label{P2}  \\
\left\{J_a,J_b\right\} &= \epsilon_{abc}J^c\,. \label{P3} 
\end{align} 

The analogous procedure can now be applied to the bracket (\ref{HDA2}). At the left hand 
side we calculate
\begin{align}
& \left\{ S[N],D[N^a] \right\} = \left\{ \Delta t P_0 + v^a K_a,\Delta x^b P_b + \varphi^b J_b \right\} \nonumber\\ 
&= \Delta t \Delta x^b \left\{ P_0,P_b \right\} + \Delta t \varphi^b \left\{ P_0,J_b \right\} \nonumber\\ 
&+ v^a \Delta x^b \left\{ K_a,P_b \right\} + v^a \varphi^b \left\{ K_a,J_b \right\}, \label{SDL}
\end{align}
while at the right hand side we have
\begin{align}
& -S\left[ N^a \partial_a N \right] = -S\left[ \Delta x^a v_a + v^a {R^a}_bx^b \right] \nonumber\\
&= \Delta x^a v_a P_0 - v^a \epsilon^{abc} \varphi_c K^b\,. \label{SDR}
\end{align}
Comparing (\ref{SDL}) with (\ref{SDR}) we obtain the brackets: 
\begin{align}
\left\{ P_0,P_a\right\} &= 0\,, \label{P4} \\
\left\{ P_0,J_a\right\} &= 0\,, \label{P5} \\
\left\{ K_a,P_b\right\} &= \delta_{ab} P_0\,, \label{P6}\\
\left\{ J_a,K_b\right\} &= \epsilon^{abc} K_c\,. \label{P7}
\end{align} 

Finally, let us consider the bracket (\ref{HDA3}). In this case the left hand side gives
\begin{align}
& \left\{ S[N_1],S[N_2] \right\} = \left\{ \Delta t_1 P_0 + v^a_1 K_a,\Delta t_2 P_0 + v^b_2 K_b \right\} \nonumber \\
&= \Delta t_1 \Delta t_2 \left\{ P_0,P_0 \right\} + (v^a_1 \Delta t_2 - v^a_1 \Delta t_1) \left\{ K_a,P_0 \right\} \nonumber \\
&+ v^a_1 v^b_2 \left\{ K_a,K_b \right\} \label{SSL}
\end{align}
and the right hand side
\begin{align}
& D\left[ s q^{ab} (N_1 \partial_b N_2 - N_2 \partial_b N_1) \right]  \nonumber \\
&= s D\left[ (v^a_1 \Delta t_1 - v^a_1 \Delta t_2) + \epsilon^{abc} (\epsilon_{bde} v_1^d v_2^e) x^c \right] \nonumber \\
&= -s (v^a_1 \Delta t_1 - v^a_1 \Delta t_2) P_a - s \epsilon_{abc} v_1^b v_2^c J^a\,, \label{SSR}
\end{align}
which leads to
\begin{align}
\left\{ K_a,P_0 \right\} &= -s P_a\,, \label{P9}\\
\left\{ K_a,K_b \right\} &= -s \epsilon^{abc} K_c\,. \label{P10}
\end{align} 
The latter brackets are the only ones which are affected by the signature $s$.

The obtained set of nine brackets (\ref{P1}-\ref{P3}), (\ref{P4}-\ref{P7}) and (\ref{P9}-\ref{P10}) defines the Poincar\'{e} (or respectively Euclidean) algebra.


\begin{thebibliography}{99}

\bibitem{Nesterov:2011gy} D.~Nesterov and S.~N.~Solodukhin, 
Nucl.\ Phys.\ B {\bf 842}, 141 (2011) 
[arXiv:1007.1246 [hep-th]].

\bibitem{Ambjorn:2005db}
  J.~Ambjorn, J.~Jurkiewicz and R.~Loll,
  Phys.\ Rev.\ Lett.\ {\bf 95}, 171301 (2005) 
  [hep-th/0505113].

\bibitem{Coumbe:2014noa}
  D.~N.~Coumbe and J.~Jurkiewicz,
  JHEP {\bf 1503}, 151 (2015) 
  [arXiv:1411.7712 [hep-th]].

\bibitem{Carlip:2009kf}
  S.~Carlip,
  AIP Conf.\ Proc.\ {\bf 1196}, 72 (2009) 
  [arXiv:0909.3329 [gr-qc]].

\bibitem{Horava:2009st} P.~Ho\v{r}ava, 
Phys.\ Rev.\ Lett.\ {\bf 102}, 161301 (2009) 
[arXiv:0902.3657 [hep-th]].

\bibitem{Lauscher:2005fy}
  O.~Lauscher and M.~Reuter,
  JHEP {\bf 0510}, 050 (2005) 
  [hep-th/0508202].

\bibitem{Calcagni:2012qn}
  G.~Calcagni,
  Phys.\ Rev.\ D {\bf 86}, 044021 (2012) 
  [arXiv:1204.2550 [hep-th]].

\bibitem{Carlip:2016dy}
S.~Carlip, 
Class.\ Quant.\ Grav.\ {\bf 32}, 232001 (2015) 
[arXiv:1506.08775 [gr-qc]].

\bibitem{Belenchia:2016ss}
A.~Belenchia, D.~M.~T.~Benincasa, A.~Marcian\`{o} and L.~Modesto, 
Phys.\ Rev.\ D {\bf 93}, 044017 (2016) 
[arXiv:1507.00330 [gr-qc]].

\bibitem{Modesto:2008jz}
  L.~Modesto,
  Class.\ Quant.\ Grav.\ {\bf 26}, 242002 (2009) 
  [arXiv:0812.2214 [gr-qc]].

\bibitem{Calcagni:2014cza}
  G.~Calcagni, D.~Oriti and J.~Th\"{u}rigen,
  Phys.\ Rev.\ D {\bf 91}, 084047 (2015) 
  [arXiv:1412.8390 [hep-th]].

\bibitem{Benedetti:2008gu}
  D.~Benedetti,
  Phys.\ Rev.\ Lett.\  {\bf 102}, 111303 (2009) 
  [arXiv:0811.1396 [hep-th]].

\bibitem{Arzano:2014jfa}
  M.~Arzano and T.~Trze\'{s}niewski,
  Phys.\ Rev.\ D {\bf 89}, 124024 (2014) 
  [arXiv:1404.4762 [hep-th]].

\bibitem{Amelino:2014pe}
G.~Amelino-Camelia, M.~Arzano, G.~Gubitosi and J.~Magueijo, 
Phys.\ Lett.\ B {\bf 736}, 317 (2014) 
[arXiv:1311.3135 [gr-qc]]

\bibitem{Bojowald:2011aa}
  M.~Bojowald and G.~M.~Paily,
  Phys.\ Rev.\ D {\bf 86}, 104018 (2012)
  [arXiv:1112.1899 [gr-qc]].

\bibitem{Cailleteau:2011kr}
  T.~Cailleteau, J.~Mielczarek, A.~Barrau and J.~Grain,
  Class.\ Quant.\ Grav.\ {\bf 29}, 095010 (2012) 
  [arXiv:1111.3535 [gr-qc]].

\bibitem{Bojowald:2015gra}
  M.~Bojowald and J.~Mielczarek,
  JCAP {\bf 1508}, 052 (2015) 
  [arXiv:1503.09154 [gr-qc]].

\bibitem{Mielczarek:2012pf}
  J.~Mielczarek,
  Springer Proc.\ Phys.\ {\bf 157}, 555 (2014) 
  [arXiv:1207.4657 [gr-qc]].

\bibitem{Bojowald:2012ux}
  M.~Bojowald and G.~M.~Paily,
  Phys.\ Rev.\ D {\bf 87}, 044044 (2013) 
  [arXiv:1212.4773 [gr-qc]].

\bibitem{Amelino-Camelia:2016gfx}
  G.~Amelino-Camelia, M.~M.~Da Silva, M.~Ronco, L.~Cesarini and O.~M.~Lecian,
  Phys.\ Rev.\ D {\bf 95}, 024028 (2017) 
  [arXiv:1605.00497 [gr-qc]].

\bibitem{Brahma:2016tsq}
  S.~Brahma, M.~Ronco, G.~Amelino-Camelia and A.~Marcian\`{o},
  Phys.\ Rev.\ D {\bf 95}, 044005 (2017) 
  [arXiv:1610.07865 [gr-qc]].

\bibitem{Ronco:2016rtp}
  M.~Ronco,
  Adv.\ High Energy Phys.\ {\bf 2016}, 9897051 (2016)
  [arXiv:1605.05979 [gr-qc]].

\bibitem{Amelino:1999ds}
G.~Amelino-Camelia, J.~Lukierski and A.~Nowicki, 
Int.\ J.\ Mod.\ Phys.\ A {\bf 14}, 4575 (1999) 
[gr-qc/9903066].

\bibitem{Mielczarek:2013rva}
  J.~Mielczarek,
  Europhys.\ Lett.\ {\bf 108}, 40003 (2014) 
  [arXiv:1304.2208 [gr-qc]].

\bibitem{Bojowald:2016hgh}
  M.~Bojowald, U.~B\"{u}y\"{u}k\c{c}am, S.~Brahma and F.~D'Ambrosio,
  Phys.\ Rev.\ D {\bf 94}, 104032 (2016)
  [arXiv:1610.08355 [gr-qc]].

\bibitem{Mielczarek:2012tn}
  J.~Mielczarek,
  AIP Conf.\ Proc.\ {\bf 1514}, 81 (2012) 
  [arXiv:1212.3527 [gr-qc]].

\bibitem{Brown:1992br}
  J.~D.~Brown and J.~W.~York, Jr.,
  Phys.\ Rev.\ D {\bf 47}, 1407 (1993)
  [gr-qc/9209012].

\bibitem{Kovacevic:2012an}
  D.~Kova\v{c}evi\'{c}, S.~Meljanac, A.~Pacho\l\ and R.~\v{S}trajn,
  Phys.\ Lett.\ B {\bf 711}, 122 (2012) 
  [arXiv:1202.3305 [hep-th]].

\bibitem{Borowiec:2010ks}
A.~Borowiec and A.~Pacho\l, 
SIGMA {\bf 6}, 086 (2010) 
[arXiv:1005.4429 [math-ph]].

\bibitem{Cianfrani:2016ss} F.~Cianfrani, J.~Kowalski-Glikman, D.~Pranzetti and G.~Rosati, 
Phys.\ Rev.\ D {\bf 94}, 084044 (2016) 
[arXiv:1606.03085 [hep-th]].

\bibitem{Magueijo:2001cr}
  J.~Magueijo and L.~Smolin,
  Phys.\ Rev.\ Lett.\ {\bf 88}, 190403 (2002) 
  [hep-th/0112090].

\bibitem{Calcagni:2013vsa}
  G.~Calcagni, A.~Eichhorn and F.~Saueressig,
  Phys.\ Rev.\ D {\bf 87}, 124028 (2013) 
  [arXiv:1304.7247 [hep-th]].

\bibitem{Lukierski:1994qt}
J.~Lukierski, H.~Ruegg and A.~Nowicki, 
J.\ Math.\ Phys.\ {\bf 35}, 2607 (1994).

\bibitem{Guedes:2013qs}
C.~Guedes, D.~Oriti and M.~Raasakka, 
J.\ Math.\ Phys.\ {\bf 54}, 083508 (2013) 
[arXiv:1301.7750 [math-ph]].

\bibitem{Freidel:2008fy}
L.~Freidel, J.~Kowalski-Glikman and S. Nowak, 
Int.\ J.\ Mod.\ Phys.\ A {\bf 23}, 2687 (2008) 
[arXiv:0706.3658 [hep-th]].

\bibitem{Isham:1975ur}
  C.~J.~Isham,
  Proc.\ Roy.\ Soc.\ Lond.\ A {\bf 351}, 209 (1976).

\bibitem{Carroll:1965} 
J.-M.~L\'{e}vy-Leblond, Annales de l'I.H.P., section A, tome {\bf 3}, no 1 (1965).

\bibitem{Bacry:1968ps} H.~Bacry and J.-M.~L\'{e}vy-Leblond, 
J.\ Math.\ Phys.\ {\bf 9}, 1605 (1968).

\bibitem{Belinsky:1970ew}
  V.~A.~Belinsky, I.~M.~Khalatnikov and E.~M.~Lifshitz,
  Adv.\ Phys.\ {\bf 19}, 525 (1970).

\bibitem{Thiemann:2008}
T.~Thiemann, ``Modern Canonical Quantum General Relativity,''
Cambridge University Press (2008).

\end{thebibliography}
\end{document}